\documentclass[prl,twocolumn,superscriptaddress,nobibnotes,letterpaper]{revtex4}

\usepackage{times}
\usepackage[pdftex]{graphicx}
\usepackage[pdftex]{epsfig}
\usepackage{amsmath}
\usepackage{pslatex}
\usepackage{amssymb}
\usepackage{amsfonts}
\usepackage{color}
\usepackage{wrapfig}
\usepackage{eucal}
\usepackage{hhline}
\usepackage{threeparttable}
\usepackage{supertabular}
\usepackage{multirow}
\usepackage{tabularx}
\usepackage[nolist,nohyperlinks]{acronym}

\def\be{\begin{equation}}
\def\ee{\end{equation}}
\def\bea{\begin{eqnarray}}
\def\eea{\end{eqnarray}}

\def\be{\begin{equation}}
\def\ee{\end{equation}}
\def\bea{\begin{eqnarray}}
\def\eea{\end{eqnarray}}

\def\kappae{\kappa_{\footnotesize\textrm{e}}}

\def\gammaom{\gamma_{\footnotesize\textrm{OM}}}
\def\gammai{\gamma_{\footnotesize\textrm{i}}}

\def\omegam{\omega_\text{m}}
\def\omegao{\omega_\text{o}}
\def\lambdao{\lambda_\text{o}}

\usepackage{hhline}
\usepackage{threeparttable}
\usepackage{supertabular}
\usepackage{multirow}
\usepackage{tabularx}

\newcolumntype{Y}{>{\centering\arraybackslash}X}

\definecolor{MyDarkBlue}{rgb}{0.1,0,0.55}
\definecolor{MyCyan}{rgb}{0.1,0.6,0.7}
\definecolor{MyGreen}{rgb}{0.1,0.6,0.1}
\definecolor{MyRed}{rgb}{0.8,0.1,0.1}
\definecolor{WGeyez}{rgb}{0.1,0.1,0.9}
\definecolor{WGoz}{rgb}{0.7,0.9,0.7}
\definecolor{WGoyez}{rgb}{0.9,0.7,0.7}
\definecolor{WGleaky}{rgb}{0.7,0.3,0.3}

\begin{document}

\pagenumbering{arabic}

\title{Two-dimensional phononic-photonic bandgap optomechanical crystal cavity}

\author{Amir H. Safavi-Naeini}
\altaffiliation[Current address: ]{ETH Z\"urich and Stanford University}
\author{Jeff T. Hill}
\altaffiliation[Current address: ]{Stanford University}
\author{Se\'an Meenehan}
\author{Jasper Chan}
\author{Simon Gr\"oblacher}
\author{Oskar Painter}
\email{opainter@caltech.edu}
\affiliation{Kavli Nanoscience Institute and Thomas J. Watson, Sr., Laboratory of Applied Physics, California Institute of Technology, Pasadena, CA 91125}
\affiliation{Institute for Quantum Information and Matter, California Institute of Technology, Pasadena, CA 91125}

\date{\today}


\begin{abstract}
We present the fabrication and characterization of an artificial crystal structure formed from a thin-film of silicon which has a full phononic bandgap for microwave $X$-band phonons and a two-dimensional pseudo-bandgap for near-infrared photons.  An engineered defect in the crystal structure is used to localize optical and mechanical resonances in the bandgap of the planar crystal.  Two-tone optical spectroscopy is used to characterize the cavity system, showing a large coupling ($g_{0}/2\pi\approx 220$~kHz) between the fundamental optical cavity resonance at $\omegao /2\pi = 195$~THz  and co-localized mechanical resonances at frequency $\omegam /2\pi \approx 9.3$~GHz.
\end{abstract}

\maketitle


Control of optical~\cite{John1987,Yablonovitch1987} and mechanical waves~\cite{Kushwaha1993,Sigalas1992} by periodic patterning of materials has been a focus of research for more than two decades. Periodically patterned dielectric media, or photonic crystals, have led to a series of scientific and technical advances in the way light can be manipulated, and has become a leading paradigm for on-chip photonic circuits~\cite{JoannopoulosJo08-book,McNab2003}.  Periodic mechanical structures, or phononic crystals, have also been developed to manipulate acoustic waves in elastic media, with myriad applications from radio-frequency filters~\cite{Olsson2009} to the control of heat flow in nanofabricated systems~\cite{Hopkins2010}.  It has also been realized that the same periodic patterning can simultaneously be used to modify the propagation of light and acoustic waves of similar wavelength~\cite{Maldovan2006,Sadat-Saleh2009}.  Such phoxonic or optomechanical crystals can be engineered to yield strong opto-acoustic interactions due to the co-localization of optical and acoustic fields~\cite{Trigo2002,Kang2009,Eichenfield2009b,Psarobas2010}.  

Utilizing silicon-on-insulator (SOI) wafers, similar to that employed to form planar photonic crystal devices~\cite{McNab2003}, patterned silicon nanobeam structures have recently been created in which strong driven interactions are manifest between localized photons in the $\lambda=1500$~nm telecom band and GHz-frequency acoustic modes~\cite{Eichenfield2009b,Safavi-Naeini2011,Chan2011}. These quasi one-dimensional (1D) optomechanical crystal (OMC) devices have led to new opto-mechanical effects, such as the demonstration of slow light and electromagnetically induced amplification~\cite{Safavi-Naeini2011}, radiation-pressure cooling of mechanical motion to its quantum ground state~\cite{Chan2011}, and coherent optical wavelength conversion~\cite{Hill2012}.  Although two-dimensional (2D) photonic crystals have been used to study localized phonons and photons~\cite{Safavi-Naeini2010a,Gavartin2011,Alegre2011}, in order to create circuit-level functionality for both optical and acoustic waves, a planar 2D crystal structure with simultaneous photonic and phononic bandgaps~\cite{Mohammadi2007,Mohammadi2010,Safavi-Naeini2010} is strongly desired.  In this Letter we demonstrate a 2D OMC structure formed from a planar ``snowflake'' crystal~\cite{Safavi-Naeini2010} which has both an in-plane pseudo-bandgap for telecom photons and a full three-dimensional bandgap for microwave $X$-band phonons.  A photonic and phononic resonant cavity is formed in the snowflake lattice by tailoring the properties of a bandgap-guided waveguide for optical and acoustic waves, and two-tone optical spectroscopy is used to characterize the strong optomechanical coupling that exists between localized cavity resonances.

\begin{figure*}[ht]
\begin{center}
\includegraphics[width=2\columnwidth]{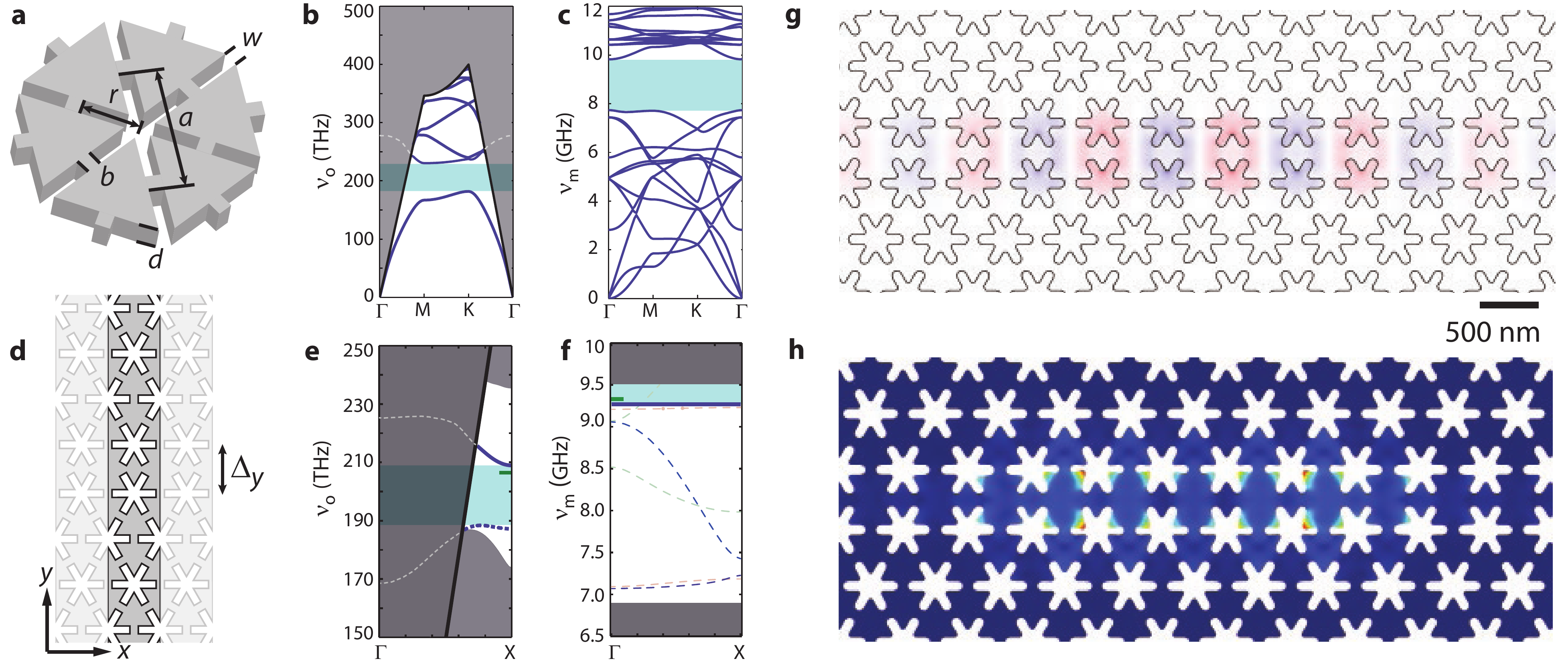}
\end{center}
\caption{\label{fig:snowflake} \textbf{a,} Snowflake crystal unit cell.  \textbf{b,} Photonic and \textbf{c,} phononic bandstructure of a silicon planar snowflake crystal with $(d,r,w,a)=(220,210,75,500)$nm.  Photonic bandstructures are computed with the MPB~\cite{MPB} mode solver and phononic bandstructures are computed with the COMSOL~\cite{COMSOL} finite-element method (FEM) solver.  In the photonic bandstructure only the fundamental even-parity optical modes (solid blue curves) of the silicon slab are shown and the grey shaded area indicates the region above the light line of the vacuum cladding.  The dashed grey curves are leaky resonances above the light line.  \textbf{d,} Unit cell schematic of a linear waveguide formed in the snowflake crystal, in which a row of snowflake holes are removed and the surrounding holes are moved inwards by $W$, yielding a waveguide width $\Delta_y = \sqrt{3}a - 2W$.  \textbf{e,} Photonic and \textbf{f,} phononic bandstructure of the linear waveguide with $(d,r,w,a,W)=(220,210,75,500,200)$nm.  Solid blue curves are waveguide bands of interest; shaded light blue regions are bandgaps of interest; green tick mark indicates the cavity mode frequencies. \textbf{g,} FEM simulated mode profile of the fundamental optical resonance at $\omegao/2\pi = 195$~THz ($\lambdao = 1530$~nm).  $E_{y}$-component of the electric field is plotted here, with red (blue) corresponding to positive (negative) field amplitude.   \textbf{h,} FEM simulated mechanical resonance displacement profile for mode with $\omega_{\text{m}}/2\pi = 9.35$~GHz and $g_0/2\pi = 250~\text{kHz}$.  Here the magnitude of the displacement is represented by color (large displacement in red, zero displacement in blue).}
\end{figure*}

The snowflake crystal~\cite{Safavi-Naeini2010}, a unit cell of which is shown in Fig.~\ref{fig:snowflake}a, is composed of a triangular lattice of holes shaped as snowflakes.  The dimensional parameters of the snowflake lattice are the radius $r$, lattice constant $a$, snowflake width $w$, and silicon slab thickness $d$. Alternatively, the structure can be thought of as an array of triangles connected to each other by thin bridges of width $b = a - 2r$. The bridge width $b$ can be used to tune the relative frequency of the low frequency acoustic-like phonon bands and the higher frequency optical-like phonon bands of the structure. For narrow bridge width, the acoustic-like bands are pulled down in frequency due to a softening of the structure for long wavelength excitations, whereas the internal resonances of the triangles that form the higher frequency optical-like phonon bands are unaffected.  This gives rise to a bandgap in the crystal, exactly analogous to phononic bandgaps in atomic crystals between their optical and acoustic phonon branches.  

As detailed in Ref.~\cite{Safavi-Naeini2010}, for the nominal lattice parameters and silicon device layer used in this work, $(d,r,w,a)=(220,210,75,500)$nm, a full three-dimensional phononic bandgap between $6.9$ and $9.5$~GHz is formed.  A corresponding pseudo-bandgap also exists for the fundamental even-parity optical guided-wave modes of the slab, extending from optical frequencies of $185$~THz to $235$~THz, or a wavelength range of $\lambda=1620$~nm to $1275$~nm. Plots of the photonic and phononic bandstructures are shown in Figs.~\ref{fig:snowflake}b and c, respectively. A significant benefit of the planar snowflake crystal is that the optical guided-wave bandgap lies substantially below the light line at the zone boundary ($\nu_{\text{ll}}\gtrsim 350$~THz), enabling low-loss guiding and trapping of light within the 2D plane.


Creation of localized defect states for phonons and photons in the quasi-2D crystal is a two-step procedure. First, a line defect is created, which acts as a linear waveguide for the propagation of optical and acoustic waves at frequencies within their respective bandgaps (see Figs.~\ref{fig:snowflake}d-f).  Second, the properties of the waveguide are modulated along its length, locally shifting the bands to frequencies that cannot propagate within the waveguide.  For the snowflake cavity studied here a small ($3\%$) quadratic variation in the radius of the snowflake holes is used to localize both the optical and acoustic waveguide modes~\cite{Safavi-Naeini2010}.  Simulated field profiles of the fundamental optical resonance ($\omegao/2\pi=195$~THz) and strongly coupled $X$-band acoustic resonance ($\omega_{\text{m}}/2\pi=9.35$~GHz) of such a snowflake crystal cavity are shown in Figs.~\ref{fig:snowflake}g and h, respectively.  Note that here we have slightly rounded features in the simulation to better approximate the properties of the crystal that is actually fabricated.  The localized acoustic mode has a theoretical optomechanical coupling of $g_0/2\pi = 250$~kHz to the co-localized optical resonance, an effective motional mass of 4 femtograms, and a zero-point-motion amplitude of $x_{\text{zpf}}=1.5$~femtometers.

\begin{figure}[t]
\begin{center}
\includegraphics[width=\columnwidth]{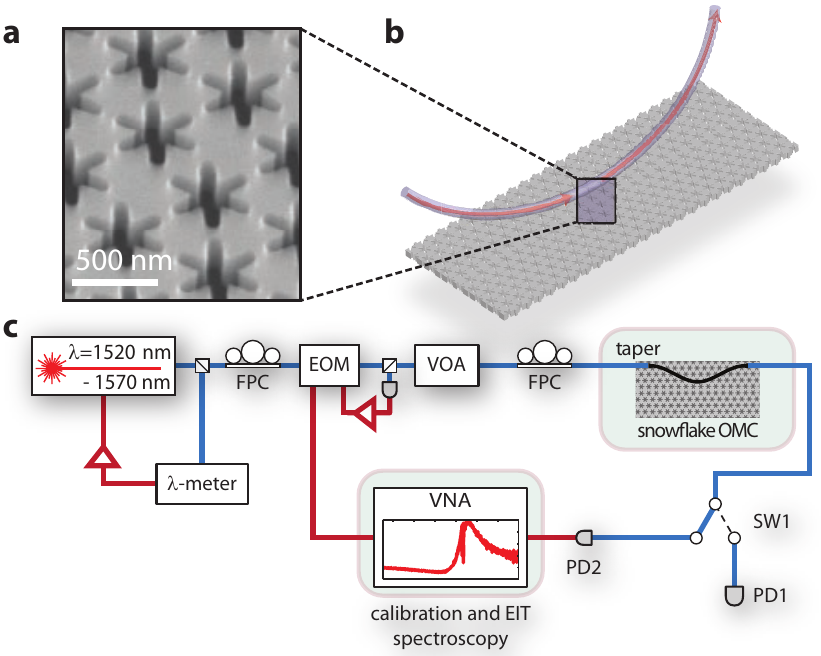}
\end{center}
\caption{\label{fig:setup} \textbf{a,} SEM image of fabricated snowflake crystal structure. \textbf{b,} Schematic showing the fiber-taper-coupling method used to optically excite and probe the snowflake cavity. \textbf{c}, Experimental setup for optical and mechanical spectroscopy of the snowflake cavity. PD $\equiv$ photodetector, VOA $\equiv$ variable optical attenuator, FPC $\equiv$ fiber polarization controller, $\lambda$-meter $\equiv$ optical wavemeter, EOM $\equiv$ electro-optic modulator, and VNA $\equiv$ vector network analyzer.}
\end{figure}

Fabrication of the snowflake OMC cavity design consists of electron beam lithography to define the snowflake pattern, a C$_4$F$_8$:SF$_6$ inductively-coupled plasma dry etch to transfer the pattern into the $220$~nm silicon device layer of an SOI chip, and a HF wet etch to remove the underlying SiO$_2$ layer to release the patterned structure.  A zoom-in of the cavity region of a fabricated device is shown in the scanning electron microscope (SEM) image of Fig.~\ref{fig:setup}a.  Testing of the fabricated devices is performed at cryogenic temperatures ($T_{\text{b}} \sim 20$~K) and high-vacuum ($P \sim 10^{-6}$~Torr) in a helium continuous-flow cryostat.  An optical taper with a localized dimple region is used to evanescently couple light into and out of individual devices with high efficiency (see Fig.~\ref{fig:setup}b).  The schematic of the full optical test set-up used to characterize the snowflake cavities is shown in Fig.~\ref{fig:setup}c and described in the figure caption.  The optical properties of the localized resonances of the snowflake cavity are determined by scanning the frequency of a narrowband tunable laser across the $\lambda=1520$-$1570$~nm wavelength band, and measuring the transmitted optical power on a photodetector.  From the normalized transmission spectrum, the resonance wavelength, the total optical cavity decay rate, and the external coupling rate to the fiber taper waveguide of the fundamental optical resonance for the device studied here are determined to be $\lambdao=1529.9$~nm, $\kappa/2\pi = 2.1~\text{GHz}$, and $\kappae/2\pi = 1.0~\text{GHz}$, respectively, corresponding to a loaded (intrinsic) optical $Q$-factor of $9.3\times 10^{4}$ ($1.8\times 10^{5}$).

\begin{figure}[t]
\begin{center}
\includegraphics[width=\columnwidth]{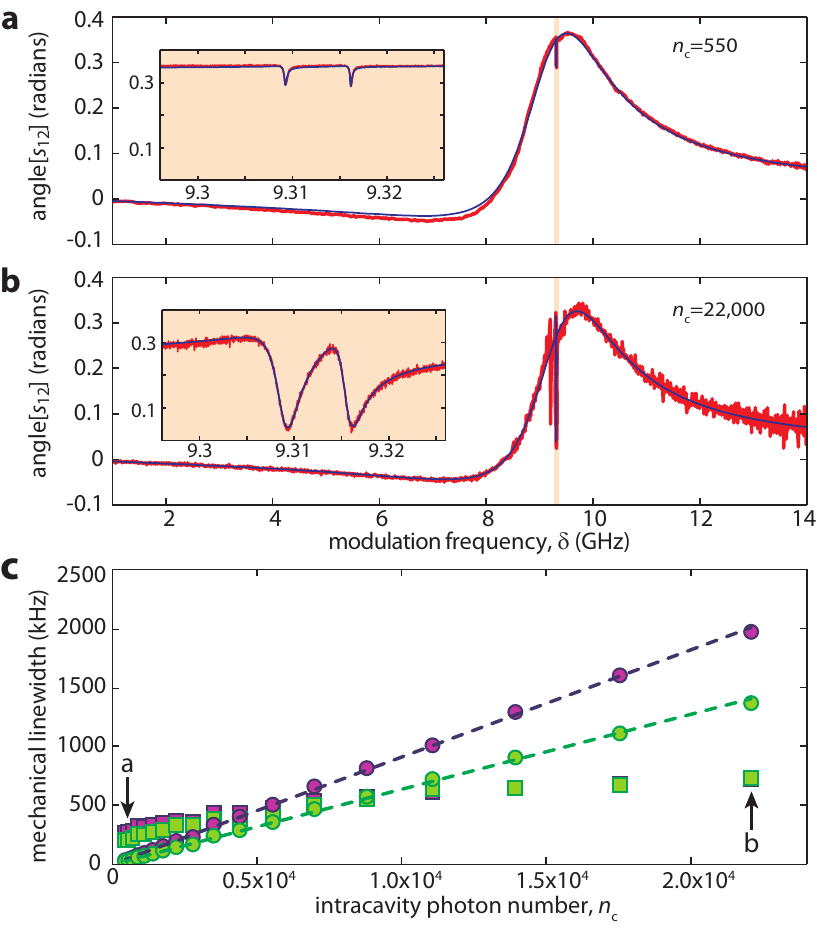}
\end{center}
\caption{\label{fig:snowflake_eit} \textbf{a,} Low ($n_{\text{c}}=550$) and \textbf{b,} high ($n_{\text{c}}=2.2\times 10^{4}$) power EIT spectra of the snowflake cavity with nominal parameters described in the text.  The insets show a zoom-in of the interference resulting from the optomechanical interaction between the optics and mechanics. The fits shown in the insets are used to extract the optomechanical coupling ($\gammaom$) and intrinsic mechanical loss rate ($\gammai$) for every optical power.  \textbf{c,} Plot of the resulting fit mechanical damping rates versus $n_c$.  $\gammai$ data are shown as squares ($\square$) and $\gammaom$ are shown as circles ($\circ$), with the low (high) frequency mode shown in green (purple).  Dashed lines correspond to linear fits to the $\gammaom$ data.}
\end{figure}

Mechanical properties of the cavity device are measured using a variant of the optical two-tone spectroscopy used to demonstrate slow-light and electromagnetically induced transparency (EIT) in optomechanical cavities~\cite{Weis2010,Safavi-Naeini2011}.  In this measurement scheme, the input laser frequency ($\omega_{\text{l}}$) is locked off-resonance from the cavity resonance ($\omegao$) at a red-detuning close to the mechanical frequency of interest, $\Delta\equiv \omegao-\omega_{\text{l}}\approx \omegam$.  Optical sideband tones are generated on the input laser beam by using an electro-optic intensity modulator driven by a microwave vector network analyzer (VNA). This modulated laser light is then sent into the cavity, and the optical cavity transmission is detected by a high speed photodetector, the output of which is connected to the input of the VNA. A sweep of the VNA modulation frequency ($\delta$) scans the upper modulated laser sideband across the optical cavity resonance, from which the $s_{12}(\delta)$ scattering parameter of the VNA yields the optomechanical response function.  

The normalized phase response (angle[$s_{12}(\delta)$]) of the snowflake cavity is shown in Figs.~\ref{fig:snowflake_eit}a and b for low and high optical input power, respectively.  Here laser power is indicated by estimated intracavity photon number, $n_{\text{c}}$, and the measured $s_{12}$ parameter is normalized by the response of the system with the laser detuned far from the cavity resonance ($> 20$~GHz).  
A zoom-in of the $s_{12}$ spectra near cavity resonance are shown in the insets to Figs.~\ref{fig:snowflake_eit}a and b, where two sharp dips are evident, corresponding to coupling to mechanical resonances of the snowflake cavity.  The frequency of the mechanical modes are in the $X$-band as expected, with $\omega_{\text{m},1}/2\pi = 9.309$~GHz and $\omega_{\text{m},2}/2\pi = 9.316$~GHz.  

In the sideband resolved, weak-coupling limit, the optomechanical coupling is given by $\gammaom \equiv 4g_0^2 n_c / \kappa$, where $g_0$ is the vacuum coupling rate and $\gammaom$ is an optically-induced damping of the mechanical resonance.  The depth of each resonance is given by the cooperativity $C = \gammaom/\gammai$, whereas the resonance width is given by $\gamma = \gammaom + \gammai$, where $\gammai$ is the intrinsic mechanical damping.  From the visibility and width of the mechanical resonance dips, $\gammaom$ and $\gammai$ are extracted and plotted versus $n_c$ in Fig.~\ref{fig:snowflake_eit}c.  The linear slope of $\gammaom$ versus $n_c$ yields a vacuum coupling rate for the higher (lower) frequency mechanical mode of $g_0/2\pi = 220$~kHz ($180$~kHz).  The intrinsic mechanical damping rate is also seen to slowly rise with optical input power, a result of parasitic optical absorption in the patterned silicon cavity structure~\cite{Chan2011}.  




\begin{figure}[t]
\begin{center}
\includegraphics[width=\columnwidth]{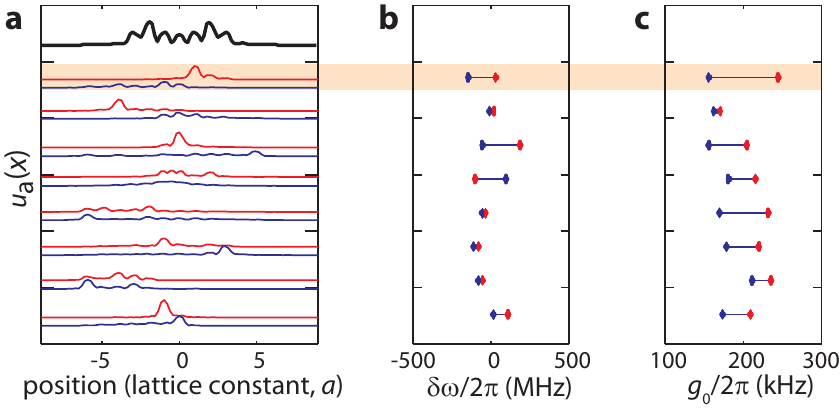}
\end{center}
\caption{\label{fig:disorder} \textbf{a,} Plot of the linear acoustic energy density profile, $u_{\text{a}}(x)$, for the localized mechanical resonances with strong optomechanical coupling to the fundamental optical resonance.  The top black curve corresponds to the unperturbed structure.  Each pair of red and blue curves correspond to the mechanical resonance with the largest and second largest magnitude of optomechanical coupling, respectively, for a different disordered structure.  $u_{\text{a}}(x)$ is computed by integrating the acoustic energy density across the transverse $y$ and $z$ dimensions.  \textbf{b,}  Plot of the corresponding mechanical frequency difference and \textbf{c,} optomechanical rate, $g_0$, for the two most strongly coupled mechanical resonances.  Here we show results for a representative 8 of the simulated disordered structures.}
\end{figure}

In order to explain the presence of two strongly coupled mechanical resonances in the measured $s_{12}$ spectrum, we note that the flat dispersion of the acoustic waveguide mode from which the cavity is formed (see Fig.~\ref{fig:snowflake}f) causes the spectrum of localized mechanical cavity modes to be highly sensitive to unavoidable fabrication disorder.  The localized optical and mechanical modes for 50 different disordered structures were calculated numerically, the results of which are summarized in Fig~\ref{fig:disorder}. Disorder was introduced into the structures by varying the width and radius of the snowflake holes in a normal distribution with 2\% standard deviation. Roughly 10$\%$ of the simulated disordered structures yielded localized mechanical resonances with frequency-splitting less than $20$~MHz and large optomechanical coupling, similar to that of the measured device. 


The snowflake 2D-OMC structure presented here provides the foundation for developing planar circuits for interacting optical and acoustic waves.  Such circuits allow for the realization of coupled arrays of devices for advanced photonic or phononic signal processing, such as the dynamic trapping and storage of optical pulses~\cite{Chang2011} or the tunable filtering and routing of microwave-over-optical signals.  In the realm of quantum optomechanics, planar 2D-OMC structures should enable operation at much lower milliKelvin temperatures, due to their improved connectivity and thermal conductance, where thermal noise is absent and quantum states of mechanical motion may be prepared and measured via quantum optical techniques.  2D-OMCs have also been theoretically proposed as the basis for quantum phononic networks~\cite{Habraken2012}, and for the exploration of quantum many-body physics in optomechanical meta-materials~\cite{Schmidt2013}.


\begin{acronym}
\acro{EIT}{electromagnetically induced transparency}
\acro{NEP}{noise equivalent power}
\acro{RPSN}{radiation pressure shot-noise}
\acro{FEM}{finite-element method}
\acro{PSD}{power spectral density}
\acro{CROW}{coupled resonator optical waveguide}
\acro{OMC}{optomechanical crystal}
\acro{EIA}{electromagnetically induced absorption}
\acro{1D}{one dimensional}
\acro{2D}{two dimensional}
\acro{3D}{three dimensional}
\acro{SOI}{silicon-on-insulator}
\acro{VNA}{vector network analyzer}

\acro{LO}{local oscillator}

\end{acronym}

\begin{acknowledgments}
The authors would like to thank T. P. M. Alegre for contributions. This work was supported by the DARPA ORCHID and MESO programs, the Institute for Quantum Information and Matter, an NSF Physics Frontiers Center with support of the Gordon and Betty Moore Foundation, and the Kavli Nanoscience Institute at Caltech. ASN and JC gratefully acknowledge support from NSERC. SG was supported by a Marie Curie International Outgoing Fellowship within the 7th European Community Framework Programme.
\end{acknowledgments}

\def\urlprefix{}
\def\url#1{}


\end{document}